\documentclass[aps,preprint]{revtex4}
\usepackage{amsmath}
\usepackage{makeidx}
\usepackage{amssymb}
\usepackage[dvipdfm]{graphicx}
\usepackage[dvipdfm]{rotating}
\usepackage{subfigure}

\begin{document}

\title{Thermofield Dynamics and Casimir Effect for Fermions}
\author{H. Queiroz$^{a}$,\thinspace\ J. C. da Silva$^{a,b}$, F. C. Khanna$
^{c,d}$, J. M. C. Malbouisson$^{a,c,}$\footnote{Corresponding author.\\
{\it E-mail addresses:}
hebe@fis.ufba.br (H. Queiroz), jcsilva@cefetba.br (J.C. da Silva),
khanna@phys.ualberta.ca (F.C. Khanna), jmalbou@phys.ualberta.ca
(J.M.C. Malbouisson), revzen@physics.technion.ac.il (M. Revzen),
asantana@fis.unb.br (A.E. Santana).}, M. Revzen$^{c,e}$, A. E.
Santana$ ^{f}$}

\affiliation{$^a$Instituto de F\'\i sica, Universidade Federal
da Bahia, Campus de Ondina, 40210-340, Salvador, BA, Brazil\\
$^b$Centro Federal de Educa\c c\~ao Tecnol\'ogica da Bahia, Rua
Em\'{\i}dio Santos, 40000-900, Salvador, BA, Brazil\\
$^{c}$Physics Department, Theoretical Physics Institute, University of
Alberta, Edmonton, AB T6G 2J1, Canada\\
$^{d}$TRIUMF, 4004, Westbrook mall, Vancouver, BC V6T 2A3,
Canada\\
$^{e}$Department of Physics, Technion - Institute of Technology,
Haifa, 32000, Israel \\
$^{f}$Instituto de F\'{\i}sica, Universidade de Bras\'{\i}lia,
70910-900, Bras\'{\i}lia, DF, Brazil}

\begin{abstract}
A generalization of the Bogoliubov transformation is developed to
describe a space compactified fermionic field. The method is the
fermionic counterpart of the formalism introduced earlier for
bosons (J. C. da Silva, A. Matos Neto, F.C. Khanna and A.E.
Santana, Phys. Rev. A 66 (2002) 052101), and is based on the
thermofield dynamics approach. We analyze the energy-momentum
tensor for the Casimir effect of a free massless fermion field in
a $d$-dimensional box at finite temperature. As a particular case
the Casimir energy and pressure for the field confined in a
3-dimensional parallelepiped box are calculated. It is found that
the attractive or repulsive nature of the Casimir pressure on
opposite faces changes depending on the relative magnitude of the
edges. We also determine the temperature at which the Casimir
pressure in a cubic box changes sign and estimate its value when
the edge of the cube is of the order of the confining lengths for
baryons.

\end{abstract}

\maketitle

\section{Introduction}

In order to calculate the partition function for a quantum field,
the Matsubara prescription is equivalent to a path-integral
calculated on \ $R^{D-1}\times S^{1}$, where $S^{1}$ is a circle
of circumference $\beta =1/T.$ This non-trivial result was
demonstrated by Polchinski \cite{pol1} at the one-loop level and
has been assumed in higher order \cite{pol2}. Under this
perspective, taking Euclidian theories, the Dolan and Jackiw
approach \cite{Jackiw} can be a useful method to perform
calculations for confined (bosonic or fermionic) fields, and
according to the Matsubara formalism the boundary conditions over
the fields are those consistent with the so called KMS (Kubo,
Martin and Schwinger) conditions, stating periodic (antiperiodic)
conditions for bosons (fermions). Such a topological result has
actually been generalized to treat different physical situations
in which fields are confined in higher space dimensions,
considering the Matsubara mechanism on a $R^{D-N}\times
S^{1_{1}}\times S^{1_{2}}...\times S^{1_{N}}$ topology, describing
time (temperature) and space confinement \cite{JMario,Ademir,GNN}.
Along the same lines, but considering the thermofield dynamics
(TFD) formalism, which is an operator formalism equivalent to the
Matsubara approach \cite{ume1,ume2,ume4,oji1,kha5,gade1,gade2} for
systems in equilibrium, a generalization of the Bogoliubov
transformation has been introduced to deal with confined boson
fields in space coordinates at finite temperature, being an
effective tool to derive different aspects of the Casimir effect
for the electromagnetic field confined between two plates
\cite{jura1}. In this context one interesting physical result is
brought about: the Casimir effect is described by a process of
condensation of the field, thus shedding a new light on the nature
of the quantum vacuum and the origin of the Casimir effect. In
this paper we extend this approach introduced for boson fields in
Ref.~\cite{jura1} to fermion fields, and we apply it to the
Casimir effect of a free massless fermionic field between two
plates, in a wave-guide and in a box.

In the realm of the quantum field theory, the Casimir effect was
first proposed as a result of the vacuum fluctuation of the
electromagnetic field confined between two plates with separation
$L$, defined by the Dirichlet boundary conditions. The effect was
an attractive force between the plates given by the negative
pressure $P=-\pi ^{2}/240L^{4}$ (we use natural units: $\hbar
=c=1$) \cite{casi1}. Over the decades, the Casimir effect has been
applied to different geometries, fields and physical conditions,
enjoying a remarkable popularity \cite
{milon,mostep,mostep3,levin,seife,boyer1,milton1,plun,bordag,ono1,car1,car2,
far1,far2,lam1,roy,rev11,tesu1} and raising interest, in
particular, in the context of microelectronics \cite{tec1,tec2} as
a practical tool for switching devices.

The effect of temperature was first studied by Lifshitz
\cite{lif,pit} who presented an alternative derivation for the
Casimir force, including an analysis of the dielectric nature of
the material between the plates. Actually, the effect of
temperature on the interaction between the conducting parallel
plates may be significant for separations greater than $3\mu m$
\cite{mehra,mostep2,mann1,mann11,mann2,mann3}. For this physical
set-up of plates, the full analysis of the thermal energy-momentum
tensor of the electromagnetic\ field was carried out by Brown and
Maclay \cite {brown}, performing the calculation of the Casimir
free-energy by using the Green's function (the local formulation)
written in a conformally invariant way
\cite{plun,robaschik,takagi}. One of our proposal here is to
derive the fermionic counterpart of the Brown and Maclay's
formula, but for a more general situation of confinement: we
consider not only the two plates, but also the case of confinement
within a $d$-dimensional box.

Casimir effect for a fermionic field is of interest in
considering, for instance, the structure of proton in particle
physics; thus its physical appeal. In particular, in the
phenomenological MIT bag model \cite{bag}, quarks are confined in
a small space region in such a way that there is no fermionic
current outside that region. The fermion field then fulfills the
so-called bag model boundary condition. The Casimir effect in such
a small region, of the order of $1.0fm$, is important to define
the process of deconfinement in heavy ion collisions at
Relativistic Heavy Ion Collider (RHIC), giving rise to the
quark-gluon plasma \cite{saito1}. The gluon field contribution for
the Casimir effect is, up to the color quantum numbers, the same
as for the electromagnetic field. For the quark field, the problem
has been often addressed by considering the case of two parallel
plates \cite {ravnd1,ravnd2,ravnd3,ravnd4,svai3,eli1}.

As first demonstrated by Johnson \cite{johns1}, for plates, the
fermionic Casimir force is attractive as in the case of the
electromagnetic field. On the other hand, depending on the
geometry of confinement, the nature of the Casimir force can
change. This is the case, for instance, for a spherical cavity and
of the Casimir-Boyer model, using mixed boundary conditions for
the electromagnetic field, such that the force is repulsive \cite
{ago1,ago11,ago2,ago3,ago4,jura2}. Therefore, the analysis
considering fermions in a wave-guide (confinement in
two-dimensions) and in a 3-dimensional box (confinement in
3-dimensions) may be of interest. We avoid here the approach based
on the sum of quantum modes, that was so important in the
calculation of the Brown and Maclay. Using, alternately, the
method developed earlier \cite{jura1}, we perform the calculation
of the non-trivial problem of the Dirac field in a finite volume
with the compactification in an $d$-dimensional box, at finite
temperature. We hope then that this possibility for calculations
may be useful in the general problem of fermions fulfilling
boundary conditions, which are some times described by coupling
with a static background field \cite{sund1}.

In order to proceed, we have to adapt the methodology of
calculations discussed earlier \cite{jura1}, to encompass a
generalized Bogoliubov transformation for fermion fields with the
MIT bag model boundary condition; equivalent to an antiperiodic
boundary condition \cite{ravnd2}. This is presented in Section 2,
where we derive the energy-momentum tensor for the fermion field
at $T\neq 0$, resulting, as an example, in the Stefan-Boltzmann
law. This is the usual calculation of TFD for fermions; however it
can also be interpreted as a confinement in the time axis, in such
a way that the field is under anti-periodic boundary conditions.
Using then this methodology for a Euclidian geometry, we can
envisage a space compactification. The usual expression of the
Casimir effect is thus calculated, considering a proper modified
Bogoliubov transformation which will describe the confinement in
the z-axis. The familiarity with this kind of calculation for
these already known results will suggest to us a general form for
a Bogoliubov transformation to describe space compactification in
arbitrary dimensions; a subject developed in Section 3. In Section
4, we consider the first application of the energy-momentum tensor
derived in Section 3: we calculate the Casimir effect considering
that the fermionic field is confined between two plates at finite
temperature. As a particular consequence, we reach the expression
derived in Ref.~\cite{ravnd1} for the Casimir energy, following
the Matsubara formalism. In Section 5 we address the problem of
the Casimir effect for a rectangular wave guide, and some
particular cases are analyzed, resulting, for instance, in an
attractive or repulsive Casimir force on opposite faces depending
on the relative magnitude of the edges of the wave-guide
transversal section. In Section 6 we consider the field confined
in a parallelepiped box, showing similar aspects as those
appearing in wage-guides. We also determine the temperature at
which the Casimir pressure in a cubic box changes sign, estimating
its value when the edge of the cube is of the order of the proton
diameter. In Section 7 some concluding remarks are presented.

\section{TFD Revisited}

The general approach to be used here can be addressed through the following
prescription, taken as a generalization of the TFD formalism \cite
{jura1,ume1,ume2} . Given an arbitrary set of operators, say $\mathcal{V}$,
with elements denoted by $A_{i},i=1,...,n$, there exists a mapping \
describing a doubling in the degrees of freedom defined by $\tau :\mathcal{V}
\rightarrow \mathcal{V}$, denoted by $\tau A\tau ^{-1}=\widetilde{A},$
satisfying the following conditions
\begin{eqnarray}
(A_{i}A_{j})\widetilde{} &=&\widetilde{A}_{i}\widetilde{A}_{j},
\label{c1til2} \\
(cA_{i}+A_{j})\widetilde{} &=&c^{\ast }\widetilde{A}_{i}+\widetilde{A}_{j},
\label{c1til3} \\
(A_{i}^{\dagger })\widetilde{} &=&(\widetilde{A}_{i})^{\dagger },
\label{c1til4} \\
(\widetilde{A}_{i})\widetilde{} &=&A_{i}.  \label{c1til5}
\end{eqnarray}
These properties are called \emph{the tilde (or dual) conjugation rules}.
The doubled Hilbert space has a new vacuum denoted by $|0,\widetilde{0}
\rangle $. Consider $\alpha \ $a $c$-number associated with macroscopic
parameters of the system such as  temperature, $\beta =1/T$ (we shall show
later that $\alpha $ may also stand for  parameters describing the space
confinement of fields as the distance between plates). The quasi-particles
in TFD are introduced by a Bogoliubov transformation given by
\begin{equation}
B(\alpha )=\left(
\begin{array}{cc}
\,\,\,\,\,u(\alpha ) & -v(\alpha ) \\
v(\alpha ) & \,\,u(\alpha )
\end{array}
\right) ,  \label{jurah3}
\end{equation}
with $\,u^{2}(\alpha )+v^{2}(\alpha )=1$.

For an arbitrary operator in $\mathcal{V}$, we use the doublet notation \cite
{ume1}
\begin{eqnarray}
(A^{a}(\alpha )) &=&\left(
\begin{array}{c}
A(\alpha ) \\
\widetilde{A}^{\dagger }(\alpha )
\end{array}
\right) =B(\alpha )\left(
\begin{array}{c}
A \\
\widetilde{A}^{\dagger }
\end{array}
\right) ,\,\,\,  \label{cs1} \\
(\,A^{a}(\alpha ))^{\dagger } &=&\left( A^{\dagger }(\alpha )\,\,,\,\,
\widetilde{A}(\alpha )\right) .\,\,  \label{cs9}
\end{eqnarray}
This notation is useful to calculate the propagator for the confined
field.

Here we are concerned with the energy-momentum tensor for a
massless fermionic field given by \cite{fermion}
\begin{eqnarray}
T^{\mu \nu }(x) &=&\langle 0|i\overline{\psi }(x^{\prime })\gamma ^{\mu
}\partial ^{\nu }\psi (x)|0\rangle |_{x^{\prime }\rightarrow x}
\label{adee1} \nonumber \\
&=&\gamma ^{\mu }\partial ^{\nu }S(x-x^{\prime })|_{x^{\prime }\rightarrow x}
\label{adee2} \nonumber \\
&=&-4i\partial ^{\mu }\partial ^{\nu }G_{0}(x-x^{\prime })|_{x^{\prime
}\rightarrow x},  \label{adee3}
\end{eqnarray}
where $S(x-x^{\prime })=-i\langle 0|T[\psi (x)\overline{\psi
}(x^{\prime })]|0\rangle $ and $G_{0}(x-x^{\prime })$ is the
propagator of the free massless bosonic field. In the
$4$-dimensional space-time (to which we shall restrict most of our
analysis) one has
\begin{eqnarray*}
G_{0}(x) &=&\frac{-1}{(2\pi )^{4}}\int d^{4}k\,\,e^{-ik\cdot x}G_{0}(k) \\
&=&\frac{-i}{(2\pi )^{2}}\frac{1}{x^{2}-i\varepsilon },
\end{eqnarray*}
where
\begin{equation*}
G_{0}(k)=\frac{1}{k^{2}+i\varepsilon }
\end{equation*}
and the Minkowski metric has the signature $(+---)$. With $T^{\mu
\nu }(x)$, we can introduce the confined ($\alpha $-dependent)
energy-momentum tensor $\mathcal{T} ^{\mu \nu (ab)}(x;\alpha )$
defined by
\begin{equation}
\mathcal{T}^{\mu \nu (ab)}(x;\alpha )=\langle T^{\mu \nu (ab)}(x;\alpha
)\rangle -\langle T^{\mu \nu (ab)}(x)\rangle ,  \label{ade122}
\end{equation}
where $T^{\mu \nu (ab)}(x;\alpha )$ is a function of the field
operators $ \psi (x;\alpha )$,$\ \widetilde{\psi }(x;\alpha )$,
according to Eq.~(\ref {cs1}), and $\langle \cdot \cdot \cdot
\rangle =\langle 0,\widetilde{0} |\cdot \cdot \cdot
|0,\widetilde{0}\rangle .$  Let us then work out this tensor.

Considering the TFD prescription \cite{ume1,sout1}, we have
\begin{equation*}
S^{(ab)}(x-x^{\prime })=\left(
\begin{array}{cc}
S(x-x^{\prime }) & 0 \\
0 & \widetilde{S}(x-x^{\prime })
\end{array}
\right) ,
\end{equation*}
with $\widetilde{S}(x-x^{\prime })=-S^{\ast }(x^{\prime }-x)$. As
a result, from Eq.~(\ref{ade122}), \ we have
\begin{equation}
\mathcal{T}^{\mu \nu (ab)}(x;\alpha )=-4i\partial ^{\mu }\partial ^{\nu
}[G_{0}^{(ab)}(x-x^{\prime };\alpha )-G_{0}^{(ab)}(x-x^{\prime
})]_{x^{\prime }\rightarrow x},  \label{jura2}
\end{equation}
corresponding to a change in Eq.~(\ref{adee2}),
$S^{(ab)}(x-x^{\prime })$ by $ S(x-x^{\prime })$. The Green's
functions in Eq.~(\ref{jura2}) are given by
\begin{equation*}
G_{0}^{(ab)}(x-x^{\prime })=\frac{-1}{(2\pi )^{4}}\int
d^{4}k\,\,G_{0}^{(ab)}(k)\,\,e^{-ik\cdot (x-x^{\prime })},
\end{equation*}
where
\begin{equation*}
G_{0}^{(ab)}(k)=\left(
\begin{array}{cc}
G_{0}(k) & 0 \\
0 & G_{0}^{\ast }(k)
\end{array}
\right) ,
\end{equation*}
and the $\alpha $-counterpart is
\begin{equation}
G_{0}^{(ab)}(x-x^{\prime };\alpha )=\frac{-1}{(2\pi )^{4}}\int
d^{4}k\,\,G_{0}^{(ab)}(k;\alpha )\,\,e^{-ik\cdot (x-x^{\prime })},
\label{jurah5}
\end{equation}
with
\begin{equation*}
G_{0}^{(ab)}(k;\alpha )=B_{k}^{-1(ac)}(\alpha
)G_{0}^{(cd)}(k)B_{k}^{(db)}(\alpha ),
\end{equation*}
where $B_{k}^{(ab)}(\alpha )$ is the Bogoliubov transformations
given in Eq.~(\ref{jurah3}). Explicitly, the components of
$G_{0}^{(ab)}(k;\alpha )$ are given by
\begin{eqnarray*}
G_0^{11}(k;\alpha ) &=&G_{0}(k)+v_{k}^{2}(\alpha )[G_{0}^{\ast
}(k)-G_{0}(k)], \\
G_0^{12}(k;\alpha ) &=&G_0^{21}(k;\alpha )=v_{k}(\alpha
)[1-v_{k}^{2}(\alpha
)]^{1/2}[G_{0}^{\ast }(k)-G_{0}(k)], \\
G_0^{22}(k;\alpha ) &=&G_{0}^{\ast }(k)+v_{k}^{2}(\alpha
)[G_{0}(k)-G_{0}^{\ast }(k)].
\end{eqnarray*}
The physical quantities are derived from the component
$G_0^{11}(k;\alpha)$. Therefore, the physical $\alpha $-tensor is
given by the component $\mathcal{ T}^{\mu \nu (11)}(x;\alpha).$

Let us consider a simple situation in which $\alpha \equiv \beta =1/T$ . In
this case $v_{k}(\beta )$ is defined through the fermion number
distribution, that is
\begin{equation*}
v_{k}(\beta )=\frac{e^{-\beta k_{0}/2}}{[1+e^{-\beta k_{0}}]^{1/2}}.
\end{equation*}
Observe that we can write
\begin{equation}
v_{k}^{2}(\beta )=\sum_{l=1}^{\infty }(-1)^{l+1}e^{-\beta k_{0}l};
\label{jurah8}
\end{equation}
leading to the thermal Green's function,
\begin{equation*}
G_{0}^{11}(k;\beta )=G_{0}(k)+\sum_{l=1}^{\infty }(-1)^{l+1}e^{-\beta
k_{0}l}[G_{0}^{\ast }(k)-G_{0}(k)].
\end{equation*}
Using this result in Eq.~(\ref{jurah5}) we derive
\begin{equation*}
G_{0}^{11}(x-x^{\prime };\beta )=G_{0}(x-x^{\prime
})+\sum_{l=1}^{\infty }(-1)^{l+1}[G_{0}^{\ast }(x^{\prime
}-x+i\beta l\widehat{n} _{0})-G_{0}(x-x^{\prime }-i\beta
l\widehat{n}_{0})],
\end{equation*}
where $\widehat{n}_{0}=(1,0,0,0)$ is a time-like vector.
Therefore, from Eq.~(\ref{jura2}), we find
\begin{equation*}
\mathcal{T}^{\mu \nu (11)}(\beta )=-4i\sum_{l=1}^{\infty
}(-1)^{l+1}\partial ^{\mu }\partial ^{\nu }[G_{0}^{\ast}(x^{\prime
}-x+i\beta l\widehat{n} _{0})-G_{0}(x-x^{\prime }-i\beta
l\widehat{n}_{0})]|_{x^{\prime }\rightarrow x}.
\end{equation*}
Performing the covariant derivatives, this expression reads
\begin{equation}
\mathcal{T}^{\mu \nu (11)}(\beta )=\frac{4}{\pi^{2}}\sum_{l=1}^{\infty
}(-1)^{l}\left[ \frac{g^{\mu \nu }-4\widehat{n}_{0}^{\mu }\widehat{n}
_{0}^{v}}{(\beta l)^{4}}\right] .  \label{jurah9}
\end{equation}

Well known results for thermal fermionic fields can be derived from this
tensor. For instance, the internal energy is given by $E(T)=\mathcal{T}
^{00(11)}(\beta ),$ that is,
\begin{equation}
E(T)=\frac{7\pi ^{2}}{60}\,T^{4},  \label{sb247}
\end{equation}
where we have used the Riemann alternating zeta-function
\begin{equation}
\varsigma (4)=\sum_{l=1}^{\infty }(-1)^{l+1}\,\frac{1}{l^{4}} =
\,\frac{7\pi ^{4}}{720}. \label{zeta4}
\end{equation}

As another application, we derive the Casimir effect at zero
temperature, by following the above calculations. For parallel
plates perpendicular to the $x^3$-direction and separated by a
distance $L$, instead of Eq.~(\ref{jurah8}), we take $\alpha
=i2L$, write
\begin{equation}
v_{k}^{2}(L)=\sum_{l=1}^{\infty }(-1)^{l+1}e^{-i2Lk_{3}l} \label{vL}
\end{equation}
and use $\widehat{n}_{3}=(0,0,0,1)$, a space-like vector. As a consequence, we
derive
\begin{equation}
\mathcal{T}^{\mu \nu (11)}(L)=\frac{4}{\pi^{2}}\sum_{l=1}^{\infty
}(-1)^{l}\left[ \frac{g^{\mu \nu }+4\widehat{n}_{3}^{\mu }\widehat{n}
_{3}^{\nu }}{(2Ll)^{4}}\right] ,  \label{jurah10}
\end{equation}
resulting in the Casimir energy and the Casimir pressure given, respectively,
by
\begin{equation}
E_{c}(L)=\mathcal{T}^{00(11)}(L)=-\frac{7\pi^{2}}{2880}
\,\frac{1}{L^{4}},
\label{sb248}
\end{equation}
\begin{equation}
P_{c}(L)=\mathcal{T}^{33(11)}(L)=-\frac{7\pi ^{2}}{960}
\,\frac{1}{L^{4}}.
\label{jurah202}
\end{equation}
Notice that the choice of $\alpha$ as a pure imaginary number is required in
order to obtain the spatial confinement, while the factor $2$ is needed to
ensure antiperiodic boundary conditions on the propagator. In the next section
we extend this procedure to the situation where multiple compactification of
(imaginary) time and spatial coordinates are simultaneously implemented.

\section{Compactification in Higher Dimensions}

In this section we calculate the Casimir effect for  massless
fermions within a $d$-dimensional (space) ``box'' at finite
temperature. For sake of generality, we will consider the
$(1+N)$-dimensional Minkowski space. We then proceed by extending
the results for the Casimir effect at zero and finite temperature
derived in the last section. Supported by these calculations, we
consider a generalization of $v(\alpha )$ (as given in
Eq.~(\ref{jurah8})) by taking $\alpha =(\alpha _{0},\alpha
_{1},\alpha _{2},...,\alpha _{N})$ and writing
\begin{eqnarray}
v_{k}^{2}(\alpha ) &=&\sum_{j=0}^{N}\sum_{l_{j}=1}^{\infty }(-1)^{l_{j}+1}\,
f(\alpha_j)\, \exp\{i\alpha_{j}l_{j}k_{j}\}  \notag \\
&&+\sum_{j<r=0}^{N}\, \sum_{l_{j},l_{r}=1}^{\infty }(-1)^{l_{j}+l_{r}+2}\,
f(\alpha_j)f(\alpha_r)\, \exp\{i\alpha_{j}l_{j}k_{j}+i\alpha_{r}l_{r}k_{r}\}
+\cdots \notag \\
&& +\sum_{l_{0},l_{1},...,l_{N}=1}^{\infty}
(-1)^{l_{0}+l_{1}+...+l_{N}+N+1}\, f(\alpha_0)f(\alpha_1)\cdots f(\alpha_N)
\exp \{i\sum_{i=0}^{N}\alpha
_{i}l_{i}k_{i}\} ,  \label{bboogg1}
\end{eqnarray}
where $f(\alpha_j)=0$ for $\alpha_j=0$ and $f(\alpha_j)=1$
otherwise. This expression leads, within the TFD formalism, to the
simultaneous compactification of any $p$ ($1\leq p \leq N+1$)
dimensions corresponding to the non null parameters $\alpha_j$,
with $\alpha_0$ corresponding to the time coordinate and
$\alpha_n$ ($n=1,...,N$) referring to the spatial ones. As we will
see later, this expression is an immediate generalization from the
results of the Casimir effect for two parallel plates at finite
temperature derived by other methods.

A more compact expression for $v_{k}^{2}(\alpha )$ is
\begin{equation}
v_{k}^{2}(\alpha) = \sum_{s=1}^{N+1}\,\sum_{\{\sigma_s \}}
\left( \prod_{n=1}^{s}f(\alpha_{\sigma_n}) \right)
\sum_{l_{\sigma_1},...,l_{\sigma_s}=1}^{\infty}
(-1)^{s+\sum_{r=1}^{s}l_{\sigma_r}}\,
\exp \{i\sum_{j=1}^{s}\alpha_{\sigma_j}l_{\sigma_j}k_{\sigma_j}\} ,
\label{boggg2}
\end{equation}
where $\{\sigma_s\}$ denotes the set of all combinations with $s$ elements,
$\{\sigma_1 ,\sigma_2 , ... \sigma_s  \}$, of the first $N+1$ natural numbers
$\{ 0,1,2, ... ,N \}$, that is all subsets containing $s$ elements, which we
choose to write in an ordered form with $\sigma_1 < \sigma_2 < \cdots <
\sigma_s$. Using this $v_{k}^{2}(\alpha )$ and generalizing the procedure
delineated in Section 2, the $(1\, 1)$-component of the
$\alpha$-dependent Green's function in momentum space becomes
\begin{eqnarray}
G_{0}^{11}(k;\alpha) & = & G_{0}(k)\, + \,
\sum_{s=1}^{N+1}\,\sum_{\{\sigma_s \}}
\left( \prod_{n=1}^{s}f(\alpha_{\sigma_n}) \right) \nonumber \\
 & & \times \sum_{l_{\sigma_1},...,l_{\sigma_s}=1}^{\infty}
(-1)^{s+\sum_{r=1}^{s}l_{\sigma_r}}\,
\exp \{i\sum_{j=1}^{s}\alpha_{\sigma_j}l_{\sigma_j}k_{\sigma_j}\}
\, [G_{0}^{\ast }(k)-G_{0}(k)]. \nonumber
\end{eqnarray}
Taking the inverse Fourier transform of this expression and defining the
vectors $\widehat{n}_{0}=(1,0,0,0,...)$, $
\widehat{n}_{1}=(0,1,0,0,...),...,$ $\widehat{n}_{N}=(0,0,0,...,1)$
in the ($1+N)$-dimensional Minkowski space, Eq.~(\ref{jura2}) leads to
\begin{eqnarray}
\mathcal{T}^{\mu \nu (11)}(\alpha )
 & = & -4i\, \sum_{s=1}^{N+1}\,\sum_{\{\sigma_s \}}
\left( \prod_{n=1}^{s}f(\alpha_{\sigma_n}) \right)
 \sum_{l_{\sigma_1},...,l_{\sigma_s}=1}^{\infty}
(-1)^{s+\sum_{r=1}^{s}l_{\sigma_r}} \nonumber \\
&&\times \,\partial^{\mu}\partial^{\nu}
\left. \left[ G_{0}^{\ast}(x^{\prime} - x +
\sum_{j=1}^{s}\xi_{\sigma_j}\alpha_{\sigma_j}l_{\sigma_j}
\widehat{n}_{\sigma_j}) - G_{0}(x - x^{\prime} -
\sum_{j=1}^{s}\xi_{\sigma_j}\alpha_{\sigma_j}l_{\sigma_j}
\widehat{n}_{\sigma_j})
\right]\right|_{x^{\prime}\rightarrow x}, \nonumber \\
 && \label{TmnGeral}
\end{eqnarray}
where $\xi_{\sigma_j}=+1$, if $\sigma_j=0$, and $\xi_{\sigma_j}=-1$
for $\sigma_j=1,2,...,N$.

As we have noticed in Section 2, to get the physical situations of
finite temperature and spatial confinement, $\alpha_0$ has to be
taken as a positive real number while $\alpha_n$, for
$n=1,2,...,N$, must be pure imaginary of the form $i2L_n$; in
these cases, one finds that $\alpha_{j}^{\ast 2} =
\alpha_{j}^{2}$. Considering such choices for the parameters
$\alpha_j$ and using the explicit form of $G_0(x)$ for the
4-dimensional space-time (corresponding to $N=3$), we obtain
\begin{eqnarray}
\mathcal{T}^{\mu \nu (11)}(\alpha ) &=& - \frac{4}{\pi^{2}}
\, \sum_{s=1}^{4}\,\sum_{\{\sigma_s \}}
\left( \prod_{n=1}^{s}f(\alpha_{\sigma_n}) \right)
 \sum_{l_{\sigma_1},...,l_{\sigma_s}=1}^{\infty}
(-1)^{s+\sum_{r=1}^{s}l_{\sigma_r}}
 \notag \\
&& \times  \frac{1}{[\sum_{j=1}^{s} \xi_{\sigma_j}(\alpha
_{\sigma_j}l_{\sigma_j})^{2}]^{2}}
\left[ g^{\mu \nu } - \frac{2\sum_{j,r=1}^{s}
(1 + \xi_{\sigma_j} \xi_{\sigma_r})(\alpha_{\sigma_j} l_{\sigma_j})
(\alpha _{\sigma_r} l_{\sigma_r})
\widehat{n}_{\sigma_j}^{\mu } \widehat{n}_{\sigma_r}^{\nu}}
{\sum_{j=1}^{s} \xi_{\sigma_j} (\alpha _{\sigma_j} l_{\sigma_j})^{2}}\right] .
\notag \\
&& \label{tan1}
\end{eqnarray}
Notice that the results given by Eqs.~(\ref{jurah9}) and
(\ref{jurah10}) are particular cases of the energy-momentum tensor
given by Eq.~(\ref{tan1}), corresponding to $\alpha=(\beta,0,0,0)$
and $\alpha=(0,0,0,i2L)$ respectively. Another important aspect is
that $\mathcal{T}^{\mu \nu (11)}(\alpha )$ is traceless, as it
should be. In order to obtain the physical meaning of $
\mathcal{T}^{\mu \nu (11)}(\alpha)$, we have to analyze particular
situations. Thus we rederive first some known (but not trivial)
results considering $N=3$. It is important to notice that
Eq.~(\ref{boggg2}) is the proper generalization of the Bogoliubov
transformation, compatible with the generalizations of the
Matsubara formalism, for the case of fermions.

\section{Casimir Effect for Two Plates}

The situation of two parallel plates at zero temperature has
already been analyzed in Section 2; for this case, taking $\alpha
= (0,0,0,i2L)$, Eq.~(\ref{tan1}) reduces to Eq.~(\ref{jurah10})
and one recovers the standard Casimir effect. Let us then consider
the case of two parallel plates at finite temperature. In this
case, both time and space confinement need to be included; this is
done by taking $\alpha = (\beta,0,0,i2L)$ in Eq.~(\ref{tan1}),
where as before $\beta^{-1}=T$ is the temperature and $L$ is the
distance between the plates perpendicular to the $x^3$-axis.
Therefore we find,
\begin{eqnarray}
\mathcal{T}^{\mu \nu (11)}(\beta ,L) &=&\frac{4}{\pi^{2}} \left\{
\sum_{l_0=1}^{\infty}(-1)^{l_0} \frac{[g^{\mu \nu} - 4
\widehat{n}_{0}^{\mu } \widehat{n}_{0}^{\nu }]}{(\beta l_0)^4} +
\sum_{l_3=1}^{\infty}(-1)^{l_3} \frac{[g^{\mu \nu} + 4
\widehat{n}_{3}^{\mu}\widehat{n}_{3}^{\nu }]}{(2 L l_3)^4}\right. \nonumber\\
 && \left. - \sum_{l_{0},l_{3}=1}^{\infty} (-1)^{l_{0}+l_{3}}
\frac{(\beta l_{0})^{2}[g^{\mu \nu} - 4\widehat{n}_{0}^{\mu }
\widehat{n}_{0}^{\nu }] + (2Ll_{3})^{2}[g^{\mu \nu} +
4\widehat{n}_{3}^{\mu } \widehat{n}_{3}^{\nu }]} {[(\beta
l_{0})^{2}+(2Ll_{3})^{2}]^{3}} \right\} .
\end{eqnarray}

Using the summation (\ref{zeta4}), the Casimir energy $E_{c}(\beta
,L)=\mathcal{T}^{00(11)}(\beta ,L)$ is given by
\begin{equation}
E_{c}(\beta ,L) = \frac{7\pi^2}{60}\,\frac{1}{\beta^4} -
\frac{7\pi^2}{2880}\,\frac{1}{L^4}
 -\,\frac{4}{\pi^{2}}\sum_{l_{0},l_{3}=1}^{\infty
}(-1)^{l_{0}+l_{3}} \frac{3(\beta l_{0})^{2}
-(2Ll_{3})^{2}}{[(\beta l_{0})^{2}+(2Ll_{3})^{2}]^{3}}.
\end{equation}
Taking the limit $L\rightarrow\infty$, this energy reduces to the
Stefan-Boltzmann term given in Eq.~(\ref{sb247}), while making
$\beta\rightarrow\infty$ one regains the Casimir effect for two
plates at zero temperature presented in Eq.~(\ref{sb248}). The
third term, which stands for the correction of temperature,
remains finite as $\beta \rightarrow 0$ and so, as expected, the
high temperature limit is dominated by the positive contribution
of the Stefan-Boltzmann term.

The Casimir pressure,
$P_{c}(\beta,L)=\mathcal{T}^{33(11)}(\beta,L)$, can be similarly
obtained as
\begin{equation}
P_{c}(\beta ,L) = \frac{7\pi^2}{180}\,\frac{1}{\beta^4} -
\frac{7\pi^2}{960}\,\frac{1}{L^4}
 +\,\frac{4}{\pi^{2}}\sum_{l_{0},l_{3}=1}^{\infty
}(-1)^{l_{0}+l_{3}} \frac{(\beta l_{0})^{2}
-3(2Ll_{3})^{2}}{[(\beta l_{0})^{2}+(2Ll_{3})^{2}]^{3}}.
\label{Pcplacas}
\end{equation}
It is to be noticed that for low temperatures (large $\beta$) the
pressure is negative but, as the temperature increases, a
transition to positive values happens. It is possible to determine
the ``critical" curve of this transition, $\beta_c=\chi_0 L$, by
searching for the value of ratio $\chi=\beta/L$ for which the
pressure vanishes; this value, $\chi_0$, is the solution of the
transcendental equation
\begin{equation*}
\frac{7\pi^2}{180}\frac{1}{\chi^{4}} - \frac{7\pi^2}{960} +
\frac{4}{\pi^2}\sum_{l,n=1}^{\infty} (-1)^{l + n} \frac{(\chi
l)^{2}-3(2 n)^{2}} {[(\chi l)^{2} + (2 n)^{2}]^{3}} = 0 ,
\label{Pzero}
\end{equation*}
given, numerically, by $\chi_0 \simeq 1.45000$.

These results, which reproduce known features about the fermionic
Casimir effect for parallel plates obtained following the
Matsubara formalism \cite{ravnd1}, suggest naturally, as we have
pointed out before, the generalization of the Bogoliubov
transformation in TFD with the expression given in
Eq.~(\ref{boggg2}). In the following sections we shall discuss
situations in which the field is confined in more than one space
directions. We start describing the Casimir effect for the
massless fermion field in a rectangular wave-guide, that is,
considering confinement in two space directions. Then we address
this problem including the effect of temperature.

\section{Casimir Effect for a Wave-guide}

The situation of a rectangular wave-guide is defined here by
considering the confinement along the $x^2$ and the $x^3$ axis.
Then the Casimir effect at zero temperature is obtained from
Eq.~(\ref{tan1}) by taking $\alpha=(0,0,i2L_{2},i2L_{3})$, that is
\begin{eqnarray}
\mathcal{T}^{\mu \nu (11)}(L_{2},L_{3}) &=&\frac{1}{4\pi^{2}} \left\{
\sum_{l_2=1}^{\infty}(-1)^{l_2} \frac{[g^{\mu \nu} + 4
\widehat{n}_{2}^{\mu } \widehat{n}_{2}^{\nu}]}{(L_2 l_2)^4} +
\sum_{l_3=1}^{\infty}(-1)^{l_3} \frac{[g^{\mu \nu} + 4
\widehat{n}_{3}^{\mu}\widehat{n}_{3}^{\nu }]}{(L_3 l_3)^4}\right. \nonumber\\
 &&  - \sum_{l_{2},l_{3}=1}^{\infty} (-1)^{l_{2}+l_{3}}\left[
\frac{(L_2 l_{2})^{2}[g^{\mu \nu} + 4\widehat{n}_{2}^{\mu }
\widehat{n}_{2}^{\nu}] + (L_3 l_{3})^{2}[g^{\mu \nu} +
4\widehat{n}_{3}^{\mu } \widehat{n}_{3}^{\nu }]}
{[(L_2 l_{2})^{2}+(L_3 l_{3})^{2}]^{3}} \right. \nonumber \\
 && \left.\left. +\,\frac{ 4(L_2 l_2)(L_3 l_3)
[\widehat{n}_{2}^{\mu}\widehat{n}_{3}^{\nu} +
\widehat{n}_{3}^{\mu} \widehat{n}_{2}^{\nu}]} {[(L_2
l_{2})^{2}+(L_3 l_{3})^{2}]^{3}}\right] \right\} .
\end{eqnarray}
Making use of Eq.~(\ref{zeta4}), the Casimir energy,
$E_{c}(L_{2},L_{3})=\mathcal{T}^{00(11)}(L_{2},L_{3}),$ is given
by
\begin{equation}
E_{c}(L_{2},L_{3})= -\,\frac{7\pi^2}{2880}\left(
\frac{1}{L_{2}^{4}} + \frac{1}{L_{3}^{4}} \right) -
\frac{1}{4\pi^2}\sum_{l_{2},l_{3}=1}^{\infty}
\frac{(-1)^{l_{2}+l_{3}}}{[(L_{2}l_{2})^{2}+(L_{3}l_{3})^{2}]^{2}},
\label{Ewg}
\end{equation}
while the Casimir pressure , $P_{c}(L_{2},L_{3})=\mathcal{T}
^{33(11)}(L_{2},L_{3})$, reads
\begin{equation}
P_{c}(L_{2},L_{3})= -\,\frac{7\pi^2}{2880}\left(
\frac{3}{L_{3}^{4}} - \frac{1}{L_{2}^{4}} \right) +
\frac{1}{4\pi^2}\sum_{l_{2},l_{3}=1}^{\infty} (-1)^{l_{2}+l_{3}}
\frac{(L_{2}l_{2})^{2}-3(L_{3}l_{3})^{2}}
{[(L_{2}l_{2})^{2}+(L_{3}l_{3})^{2}]^{3}} . \label{Pwg}
\end{equation}

For a square wave-guide ($L_2=L_3=L$), the Casimir energy and the
Casimir pressure (in this case, $\mathcal{T}^{33(11)}=\mathcal{T}
^{22(11)}$) reduce to
\begin{equation}
E_{c}(L) = - \left( \frac{7\pi^2}{1440} + \frac{{\cal
C}_2}{4\pi^2} \right) \frac{1}{L^4} ,
\end{equation}
\begin{equation}
P_{c}(L) = - \left( \frac{7\pi^2}{1440} + \frac{{\cal
C}_2}{4\pi^2} \right) \frac{1}{L^4} ,
\end{equation}
where the constant ${\cal C}_2$ is given by
\begin{equation*}
{\cal C}_2 = \sum_{l,n=1}^{\infty}\frac{(-1)^{l+n}}{(l^2 + n^2)^2}
\simeq 0.19368 .
\end{equation*}
We see that $E_c$ and $P_c$ for a square wave-guide behave, as
functions of $L$, in the same way as in the case of two parallel
plates, both being negative, but with the energy decreasing
(increasing in absolute value) and the pressure increasing
(smaller absolute value) in the wave-guide case as compared with
the two plates situation.

\begin{figure}[h]
\centering
\includegraphics[{height=6.0cm,width=8.0cm}]{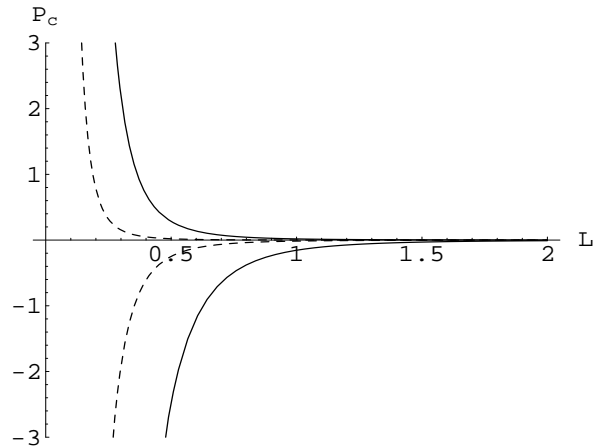}
\caption{The Casimir pressure, $P_{c}=\mathcal{T}^{33(11)}$, for
some rectangular wave-guides, as a function of $L$ ($=L_2$): the
full and the dashed lines below the horizontal axis correspond to
$\xi=0.8$ and $\xi=1.2$, respectively; the dashed and the full
lines above the horizontal axis refer to $\xi=1.4$ and $\xi=2.0$,
respectively.} \label{Fig1}
\end{figure}

On the other hand, if the case $L_2 \neq L_3$ is considered,
although $E_c$ remains negative whatever the ratio $\xi=L_3/L_2$
is, it is clear from Eq.~(\ref{Pwg}) that the sign of the Casimir
pressure depends heavily on the relative magnitude of $L_2$ and
$L_3$. In fact, a transition from negative to positive pressure is
observed as $\xi$ is increased; this feature is presented in
Figure \ref{Fig1} where we plot $P_{c}=\mathcal{T}^{33(11)}$ for
some rectangular wave-guides (characterized by different values of
$\xi$) as a function of $L$ ($=L_2$). These plots indicate the
existence of a specific value of the ratio $\xi$, $\xi_0$, for
which the Casimir pressure vanishes identically. This value is the
solution of the transcendental equation
\begin{equation*}
-\,\frac{7\pi^2}{2880}\left( \frac{3}{\xi^{4}} - 1 \right) +
\frac{1}{4\pi^2}\sum_{l,n=1}^{\infty} (-1)^{l + n}
\frac{l^{2}-3(\xi n)^{2}} {[l^{2} + (\xi n)^{2}]^{3}} = 0 ,
\label{Pzero}
\end{equation*}
which is given, numerically, by $\xi_0 \simeq 1.37955$; all
rectangular wave-guides with the ratio between $L_3$ and $L_2$
equal to $\xi_0$ have null Casimir pressure $P_c=\mathcal{T}
^{33(11)}$.

It is important to notice that the same kind of reasoning applies
to $\mathcal{T} ^{22(11)}(L_{2},L_{3})$, which can be obtained
from $\mathcal{T}^{33(11)}$ by exchanging $L_2 \leftrightarrow
L_3$. So the force on the faces of the wage-guide perpendicular to
the $x^2$-direction will also change from attractive to repulsive
if the ratio $L_2/L_3=\xi^{-1}$ increases, passing by the value
$\xi_0$. One sees that $\mathcal{T}^{33(11)}$ and
$\mathcal{T}^{22(11)}$ will never be simultaneously positive; in
fact, both are negative for $\xi_0^{-1} < \xi < \xi_0$ but they
have opposite signs whenever $\xi > \xi_0$ or  $\xi < \xi_0^{-1}$.
It is also worth mentioning that, in a sense, this transition is
similar to the transition from negative to positive pressures for
two plates as the temperature is increased; both come from the
compactification of a second space-time coordinate.

We now incorporate the effect of temperature in the wage-guide. To do so, we
consider Eq.~(\ref{tan1}) with $\alpha=(\beta,0,i2L_{2},i2L_{3})$. Then the
Casimir energy becomes
\begin{eqnarray}
E_{c}(\beta,L_{2},L_{3}) &=& \frac{7\pi^2}{60}\,\frac{1}{\beta^4} -
\,\frac{7\pi^2}{2880}\left(\frac{1}{L_{2}^{4}} + \frac{1}{L_{3}^{4}} \right)
+ \frac{4}{\pi^2}\sum_{l_{0},l_{2}=1}^{\infty}(-1)^{l_{0}+l_{2}}
\frac{3(\beta l_{0})^{2}-(2 L_{2}l_{2})^{2}}
{[(\beta l_{0})^{2}+(2 L_{2}l_{2})^{2}]^{3}} \nonumber \\
 && + \frac{4}{\pi^2}\sum_{l_{0},l_{3}=1}^{\infty}(-1)^{l_{0}+l_{3}}
\frac{3(\beta l_{0})^{2}-(2 L_{3}l_{3})^{2}}
{[(\beta l_{0})^{2}+(2 L_{3}l_{3})^{2}]^{3}}
- \frac{1}{4\pi^2}\sum_{l_{2},l_{3}=1}^{\infty}
\frac{(-1)^{l_{2}+l_{3}}}{[(L_{2}l_{2})^{2}+(L_{3}l_{3})^{2}]^{2}} \nonumber \\
 && - \frac{4}{\pi^2}\sum_{l_{0},l_{2},l_{3}=1}^{\infty}
(-1)^{l_{0}+l_{2}+l_{3}}
\frac{3(\beta l_{0})^{2}-(2 L_{2}l_{2})^{2}-(2 L_{3}l_{3})^{2}}
{[(\beta l_{0})^{2}+(2 L_{2}l_{2})^{2}+(2 L_{3}l_{3})^{2}]^{3}} ,
\label{EwgT}
\end{eqnarray}
while the Casimir pressure, $P_{c}=\mathcal{T}^{33(11)}$, is given
by
\begin{eqnarray}
P_{c}(\beta,L_{2},L_{3}) &=& \frac{7\pi^2}{180}\,\frac{1}{\beta^4}
-\,\frac{7\pi^2}{2880}\left(\frac{3}{L_{3}^{4}} - \frac{1}{L_{2}^{4}} \right)
+\,\frac{4}{\pi^2}\sum_{l_{0},l_{3}=1}^{\infty}(-1)^{l_{0}+l_{3}}
\frac{(\beta l_{0})^{2}-3(2 L_{3}l_{3})^{2}}
{[(\beta l_{0})^{2}+(2 L_{3}l_{3})^{2}]^{3}}
\nonumber \\
 && +\, \frac{4}{\pi^2} \sum_{l_{0},l_{2}=1}^{\infty}
\frac{(-1)^{l_{0}+l_{2}}}{[(\beta l_{0})^{2}+(2 L_{2}l_{2})^{2}]^{2}}
+\, \frac{1}{4\pi^2}\sum_{l_{2},l_{3}=1}^{\infty} (-1)^{l_{2}+l_{3}}
\frac{(L_{2}l_{2})^{2}-3(L_{3}l_{3})^{2}}
{[(L_{2}l_{2})^{2}+(L_{3}l_{3})^{2}]^{3}} \nonumber \\
&& - \frac{4}{\pi^2}\sum_{l_{0},l_{2},l_{3}=1}^{\infty}
(-1)^{l_{0}+l_{2}+l_{3}}
\frac{(\beta l_{0})^{2}+(2 L_{2}l_{2})^{2}-3(2 L_{3}l_{3})^{2}}
{[(\beta l_{0})^{2}+(2 L_{2}l_{2})^{2}+(2 L_{3}l_{3})^{2}]^{3}}. \label{PwgT}
\end{eqnarray}

For simplicity we concentrate on the problem of a square
wave-guide at finite temperature. As in the two-plates case, the
Casimir energy passes from negative to positive values with
increasing temperature, as expected since the Stefan-Boltzmann
term dominates all others as $\beta \rightarrow 0$. Let us then
look at the Casimir pressure $P_{c}=\mathcal{T}^{33(11)}$
($=\mathcal{T}^{22(11)}$). Taking $L_2=L_3=L$ and defining
$\chi=\beta/L$, Eq.~(\ref{PwgT}) can be written as
\begin{equation}
P_c(\chi,L) = g(\chi)\,\frac{1}{L^4} , \label{PxiL}
\end{equation}
where
\begin{eqnarray}
g(\chi) & = & \frac{7\pi^2}{180}\,\frac{1}{\chi^4} -
\,\frac{7\pi^2}{1440} -\,\frac{{\cal C}_2}{2\pi^2} +
\,\frac{4}{\pi^2}\sum_{l,n=1}^{\infty}(-1)^{l+n} \,\frac{(\chi
l)^{2}-3(2 n)^{2}} {[(\chi l)^{2}+(2 n)^{2}]^{3}} \nonumber \\
 && + \,\frac{4}{\pi^2}\sum_{l,n=1}^{\infty} \frac{(-1)^{l+n}}{[(\chi
l)^{2}+(2 n)^{2}]^{2}}- \frac{4}{\pi^2}\sum_{l,n,r=1}^{\infty}
(-1)^{l+n+r} \frac{(\chi l)^{2}+(2 n)^{2}-3(2 r)^{2}} {[(\chi
l)^{2}+(2 n)^{2}+(2 r)^{2}]^{3}}. \label{gchi}
\end{eqnarray}
As in the case of two plates, the Casimir pressure in a square
wage-guide passes from negative to positive values as the
temperature increases, the transition point given by the value of
$\chi$, $\chi_0$, such that $g(\chi_0)=0$. One finds (numerically)
$\chi_0 \simeq 1.60224$, and so write down the critical curve
$\beta_c = \chi_0\, L$. In the general case, $L_2 \neq
L_3$, increasing the temperature tends to make all diagonal
components of $\mathcal{T}^{\mu\nu(11)}$ positive.

In the next section we analyze the Casimir effect for a massless
fermion field confined in a rectangular parallelepiped box, taking
into account the effect of temperature.

\section{Casimir Effect for a box}

We now consider the field confined in a $3$-dimensional closed box
having the form of a rectangular parallelepiped with faces $L_1$,
$L_2$ and $L_3$. At zero temperature, the physical energy-momentum
tensor is obtained from Eq.~(\ref{tan1}) by taking
$\alpha=(0,i2L_1,i2L_{2},i2L_{3})$. The Casimir energy is then
given by
\begin{eqnarray}
E_c(L_1,L_2,L_3)&=& -\,\frac{7\pi^2}{2880}\left(
\frac{1}{L_{1}^{4}} + \frac{1}{L_{2}^{4}} + \frac{1}{L_{3}^{4}}
\right) - \,\frac{1}{4\pi^2}\sum_{l_{1},l_{2}=1}^{\infty}
\frac{(-1)^{l_{1}+l_{2}}}{[(L_{1}l_{1})^{2}+(L_{2}l_{2})^{2}]^{2}}
 \nonumber \\
 && -\,\frac{1}{4\pi^2}\sum_{l_{1},l_{3}=1}^{\infty}
\frac{(-1)^{l_{1}+l_{3}}}{[(L_{1}l_{1})^{2}+(L_{3}l_{3})^{2}]^{2}}
-,\frac{1}{4\pi^2}\sum_{l_{2},l_{3}=1}^{\infty}
\frac{(-1)^{l_{2}+l_{3}}}{[(L_{2}l_{2})^{2}+(L_{3}l_{3})^{2}]^{2}}
  \nonumber \\
 && +\,\frac{1}{4\pi^2}\sum_{l_{1},l_{2},l_{3}=1}^{\infty}
\frac{(-1)^{l_{1}+l_{2}+l_{3}}}{[(L_{1}l_{1})^{2}+(L_{2}l_{2})^{2}+
(L_{3}l_{3})^{2}]^{2}} , \label{Ebox}
\end{eqnarray}
and the Casimir pressure, $P_{c}=\mathcal{T}^{33(11)}$, reads
\begin{eqnarray}
P_c(L_1,L_2,L_3)&=&
-\,\frac{7\pi^2}{2880}\left(\frac{3}{L_{3}^{4}} -
\frac{1}{L_{1}^{4}} - \frac{1}{L_{2}^{4}} \right)
+\,\frac{1}{4\pi^2}\sum_{l_{1},l_{3}=1}^{\infty}
(-1)^{l_{1}+l_{3}} \frac{(L_{1}l_{1})^{2}-3(L_{3}l_{3})^{2}}
{[(L_{1}l_{1})^{2}+(L_{3}l_{3})^{2}]^{3}} \nonumber \\
 && + \,\frac{1}{4\pi^2}\sum_{l_{1},l_{2}=1}^{\infty}
\frac{(-1)^{l_{1}+l_{2}}}{[(L_{1}l_{1})^{2}+(L_{2}l_{2})^{2}]^{2}}
+\,\frac{1}{4\pi^2}\sum_{l_{2},l_{3}=1}^{\infty}
(-1)^{l_{2}+l_{3}} \frac{(L_{2}l_{2})^{2}-3(L_{3}l_{3})^{2}}
{[(L_{2}l_{2})^{2}+(L_{3}l_{3})^{2}]^{3}} \nonumber \\
&& - \,\frac{1}{4\pi^2}\sum_{l_{1},l_{2},l_{3}=1}^{\infty}
(-1)^{l_{1}+l_{2}+l_{3}}
\frac{(L_{1}l_{1})^{2}+(L_{2}l_{2})^{2}-3(L_{3}l_{3})^{2}}
{[(L_{1}l_{1})^{2}+(L_{2}l_{2})^{2}+(L_{3}l_{3})^{2}]^{3}} .
\label{Pbox}
\end{eqnarray}

\begin{figure}[h]
\centering
\includegraphics[{height=6.0cm,width=8.0cm}]{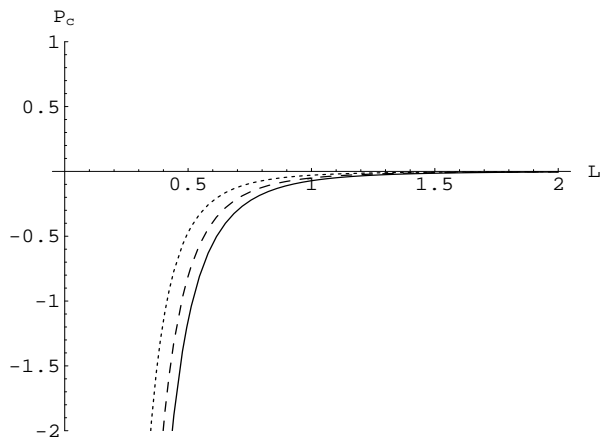}
\caption{The Casimir pressure, $P_{c}$, for two parallel plates
separated by a distance $L$ (full line); for a square wave-guide
with transversal section of edge $L$ (dashed line); and for a
cubic box of edge $L$ (dotted line).} \label{Fig2}
\end{figure}

For a cubic box ($L_1=L_2=L_3=L$), the Casimir energy and the
Casimir pressure (now one has $\mathcal{T}^{33(11)} =
\mathcal{T}^{22(11)} = \mathcal{T}^{11(11)}$) become
\begin{equation}
E_{c}(L) = - \left( \frac{7\pi^2}{960} + \frac{3{\cal C}_2 - {\cal
C}_3}{4\pi^2} \right) \frac{1}{L^4} ,
\end{equation}
\begin{equation}
P_{c}(L) = - \left( \frac{7\pi^2}{2880} + \frac{3{\cal C}_2 -
{\cal C}_3}{12\pi^2} \right) \frac{1}{L^4} ,
\end{equation}
where the constant ${\cal C}_3$ is given by
\begin{equation*}
{\cal C}_3 = \sum_{l,n,r=1}^{\infty}\frac{(-1)^{l+n+r}}{(l^2 + n^2
+ r^2)^2} \simeq -0.06314 .
\end{equation*}
One sees that both energy and pressure in cubic boxes behave
similarly to the cases of two parallel plates and of square
wave-guides; in Figure \ref{Fig2} we plot the Casimir pressure for
all these symmetrical situations, for comparison. It is curious
that, in the units we are using, the Casimir pressure is three
times the energy for parallel plates, they are equal in a square
wave-guide, while in a cubic box one has $P_c(L)=E_c(L)/3$.

Changing the relative magnitude of the edges of the parallelepiped
box leads to similar effects as in the case of the rectangular
wave-guide. For example, taking $L_1=L_2=L$ and defining
$\xi=L_3/L$, one can show that
$P_{c}(L,\xi)=\mathcal{T}^{33(11)}(L,\xi)$ vanishes for
$\xi=\xi_0\simeq 1.15274$, being negative for $\xi < \xi_0$ and
positive for $\xi > \xi_0$. However, as in the case of the
wave-guide, $\mathcal{T}^{22(11)}$ ($=\mathcal{T}^{11(11)}$ in the
present situation) will be negative whenever $\mathcal{T}^{33(11)}
> 0$.

\begin{figure}[h]
\centering
\includegraphics[{height=6.0cm,width=8.0cm}]{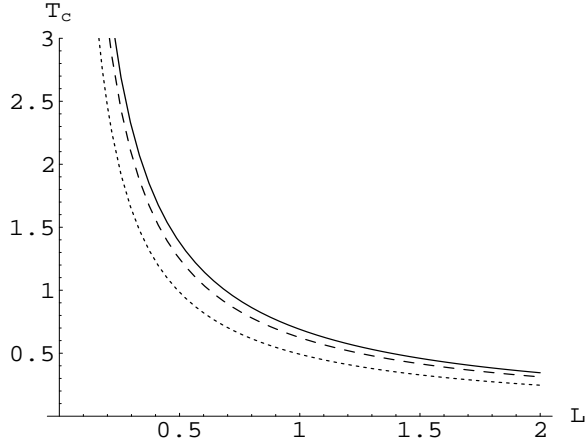}
\caption{Critical curves for the transitions from negative to
positive Casimir pressure, induced by the temperature, for: (i)
two parallel plates separated by a distance $L$ (full line); (ii)
square wave-guide of transversal section of edge $L$ (dashed
line); (iii) cubic box of edge $L$ (dotted line). In all cases,
the points below the curves correspond to $P_c<0$, while above
them one has $P_c>0$.} \label{FigT}
\end{figure}

To treat the effect of temperature in the case of a box, all four
coordinates in the Minkowski space have to be compactified by
considering $\alpha = (\beta,i2L_1,12L_2,i2L_3)$ in
Eq.~(\ref{tan1}). This amounts to the addition to
Eqs.~(\ref{Ebox}) and (\ref{Pbox}) terms involving $\beta$ and the
distances $L_j$ like those appearing in Eqs.~(\ref{EwgT}) and
(\ref{PwgT}). In the simpler case of a cubic box, one finds that
the Casimir pressure changes from negative to positive values when
the ratio $\chi=\beta/L$ passes through the value $\chi_0 \simeq
2.0323$. The critical curves, $\beta_c = \chi_0 L$, for all
symmetrical cases analyzed here (parallel plates, square
wave-guide and cubic box) appear in Figure \ref{FigT}. One notices
that the behaviors of $T_c \times L$, in all three cases, are very
similar one to each other, with $T_c$ scaling with the inverse
fourth power of the relevant length $L$.

On the basis of these results, we can estimate the temperature at
which the Casimir pressure in a cubic box changes from negative to
positive values considering the edge of the cube of the order of
confining lengths for hadronic matter. Such an estimative provides
a crude idea of the importance of Casimir effect in the
deconfinement transition for hadrons. Taking $L \approx 1 {\rm
fm}$, a length of the order of hadron radius, one finds $T_c =
(\chi_0 L)^{-1}\approx 100 \,{\rm MeV}$ from quark degrees of
freedom considering only one flavor quark. For a real hadron there
are two flavors, $u$ and $d$, each with 3 colors and an octet of
gluons. In order to calculate the Casimir energy and pressure, the
results for bosonic and fermionic fields have to be combined.

\section{Concluding Remarks}

In this paper we have presented a generalization of the Bogoliubov
transformation in order to describe a massless fermion field
compactified in a $d$-dimensional box at finite temperature. We
write the (traceless) energy-momentum tensor from which we
calculate and compare explicit expressions for the Casimir
pressure and the Casimir energy, corresponding to different cases
of confinement. The Casimir pressure at zero temperature for the
cases of two parallel plates, square wave-guides and cubic boxes
are negative (see Figure \ref{Fig2}), imposing then an attractive
force on all faces of the volume confining the fermionic field. On
the other hand, for rectangular wave-guides and parallelepiped
boxes the pressure on opposite faces can be made positive,
representing a repulsive force, by appropriately choosing the
ratios between the edges of the rectangle and between the edges of
the cube. However, at zero temperature, one cannot make positive
the pressure for all opposite faces of the wave-guide or of the
box simultaneously; this can be achieved only by raising the
temperature sufficiently. We have calculated the Casimir pressure
at finite temperature for the cases of two parallel plates, square
wave-guides and cubic boxes and determined the critical curves for
these transitions from negative to positive pressures (see Figure
\ref{FigT}). Since repulsive Casimir force can have a direct
influence on the description of quark deconfinement, we have
estimated this critical temperature for a cubic box of edge equal
to $1 {\rm fm}$ (which is the order of magnitude of nuclear
dimensions) obtaining $T_c \approx 100\,{\rm MeV}$. It is
important to emphasize that these results are only an estimate
since these are based on the presence of one fermion flavor field
in a cubic box. For realistic calculations, a spherical geometry
with full account of the color and flavor degrees of freedom has
to be considered.

It is important to stress the simplicity of these calculations,
that is clear by comparing with the known results for the fermion
Casimir effect; see for instance \cite{milton1,saito1,ravnd1}. We
have avoided usual procedures, the intricate method based on the
sum of the quantum modes of the fields, satisfying some given
boundary conditions. Indeed, instead of the sum of modes, we have
used the generalized Bogoliubov transformation to define the
Casimir effect which provides an elegant physical interpretation
of the effect as a consequence of the condensation in the vacuum
for the fermion field (in a similar fashion as was carried out for
the case of bosons \cite {jura1}). Taking advantages of these
practical proposals, the method developed here can be useful for
calculations involving other geometries with spherical or
cylindrical symmetries. This analysis will be developed in more
detail elsewhere.

\begin{description}
\item \textbf{Acknowledgments:} This work was supported by CNPQ of
Brazil, by NSERC of Canada, by the Technion-Haifa University Joint
Research Fund, and by the Fund for Promotion of Sponsored Research
at the Technion, Israel. One of us (AES) thanks M. Amato and T. M.
Rocha-Filho for  stimulating discussions and their interest in
this work. One of us (JMCM) thanks the Theoretical Physics
Institute and Vice-President (Research) for partial support during
his stay at the University of Alberta.
\end{description}


\begin{thebibliography}{99}
\bibitem{pol1} J. Polchinski, Commun. Math. Phys. \textbf{104} (1986) 37.

\bibitem{pol2} J. J. Atick and E. Witten, Nucl. Phys. B \textbf{310} (1988)
291.

\bibitem{Jackiw} L. Dolan and R. Jackiw, Phys. Rev. D \textbf{9} (1974)
3320.

\bibitem{JMario} A. P. C. Malbouisson and J. M. C. Malbouisson, J. Phys. A:
Math. Gen. \textbf{35} (2002) 2263.

\bibitem{Ademir} A. P. C. Malbouisson, J. M. C. Malbouisson and A. E.
Santana, Nucl. Phys. B \textbf{631} (2002) 83.

\bibitem{GNN} A. P. C. Malbouisson, J. M. C. Malbouisson, A. E. Santana, J.
C. Silva , Phys. Lett. B \textbf{583} (2004) 373.

\bibitem{ume1} H. Umezawa, \emph{Advanced Field Theory: Micro, Macro and
Thermal Physics}\textit{\ }(AIP, New York, 1993).

\bibitem{ume2} Y. Takahashi and H. Umezawa, Coll. Phenomena \textbf{2}
(1975) 55 (Reprinted in Int. J. Mod. Phys. 10 (1996) 1755).

\bibitem{ume4} H. Umezawa, H. Matsumoto and M. Tachiki,\emph{\ Thermofield
Dynamics and Condensed States} (North-Holland, Amsterdan, 1982).

\bibitem{oji1} I. Ojima, Ann. Phys. (N. Y.) \textbf{137 }(1981) 1.

\bibitem{kha5} A. E. Santana, A. Matos Neto, J. D. M. Vianna and F. C.
Khanna, Physica A \textbf{280} (2000) 405.

\bibitem{gade1} M. C. B. Abdala, A. L. Gadelha and I. V. Vancea, Phys. Rev.
D \textbf{64} (2001) 086005.

\bibitem{gade2} M. C. B. Abdala, A. L. Gadelha and I. V. Vancea, Int. J.
Mod. Phys. A \textbf{18} (2003) 2109.

\bibitem{jura1} J. C. da Silva, A. Matos Neto, F.C. Khanna and A. E.
Santana, Phys. Rev. A \textbf{66} (2002) 052101.

\bibitem{casi1} H. B. G. Casimir, Proc. Ned. Akad. Wet. B \textbf{51} (1948)
793.

\bibitem{milon} P. W. Milonni, \emph{The Quantum Vacuum} (Academic, Boston,
1993).

\bibitem{mostep} V. M. Mostepanenko and N.N. Trunov, \emph{The Casimir
Effect and its Applications} (Clarendon, Oxford, 1997).

\bibitem{mostep3} M. Bordag, U. Mohideed and V. M. Mostepanenko, \emph{New
Developments in Casimir Effect}, quant-ph/0106045, Phys. Rep.
\textbf{353} (2001) 1.

\bibitem{levin} F. S. Levin and D.A. Micha (Eds.), \emph{Long Range Casimir
Forces} (Plenum, New York, 1993).

\bibitem{seife} C. Seife, Science \textbf{275} (1997) 158.

\bibitem{boyer1} T. H. Boyer, Am. J. Phys. \textbf{71} (2003) 990.

\bibitem{milton1} K. A. Milton, \emph{The Casimir Effect: Physical
Manifestations of Zero Point Energy}, hep-th/9901011.

\bibitem{plun} G. Plunien, B. M\"{u}eller and W. Greiner, Phys. Rep. \textbf{
134 }(1986) 87.

\bibitem{bordag} M. Bordag (Ed.), \emph{The Casimir Effect 50 Years Later}
(World Scientific, Singapore, 1999).

\bibitem{ono1} G. Bressi, G. Carugno, R. Onofrio and G. Ruoso, Phys. Rev. Lett.  \textbf{88} (2002) 041804.

\bibitem{car1} F. Caruso, N. P. Neto, B. F. Svaiter and N. F. Svaiter, Phys.
Rev. D \textbf{43 }(1991) 1300.

\bibitem{car2} F. Caruso, R. de Paola and N. F. Svaiter, Int. J. Mod. Phys.
A \textbf{14 }(1999) 2077.

\bibitem{far1} M. V. Cougo-Pinto, C. Farina and A. Ten\'{o}rio, Braz. J.
Phys. \textbf{29 }(1999) 371.

\bibitem{far2} D. T. Alves, C. Farina and A. C. Tort, Phys. Rev. A \textbf{63%
} (2000) 4102.

\bibitem{lam1} S. K. Lamoreaux, Am. J. Phys.\textbf{\ 67} (1999) 850.

\bibitem{roy} U. Mohideen and A. Roy, Phys. Rev. Lett. \textbf{81} (1998)
4549.

\bibitem{rev11} O. Kenneth, I. Klich, A. Mann and M. Revzen, Phys. Rev.
Lett. \textbf{89} (2002) 033001.

\bibitem{tesu1} T. Maruyama, K. Tsushima and A. Faessler, Nucl. Phys. A
\textbf{537 }(1992) 303.

\bibitem{tec1} F. Serry, D. Walliser and G. J. Maclay, J. Appl. Phys.
\textbf{84 }(1998) 2501.

\bibitem{tec2} E. Buks and M. L. Roukes, Phys. Rev. B \textbf{63 }(2001)
033402.

\bibitem{lif} E. M. Lifshitz, Sov. Phys. JETP \textbf{\ 2} (1956) 73.

\bibitem{pit} I. E. Dzyaloshinskii, E. M. Lifshitz and L. P. Pitaevskii,
Adv. Phys.\textbf{10 }(1961) 165.

\bibitem{mehra} J. Mehra, Physica, \textbf{37} (1967) 145. 

\bibitem{mostep2} G. L. Klimchitskaya and V. M. Mostepanenko, Phys. Rev. A
\textbf{63 }(2001) 062108.

\bibitem{mann1} M. Revzen, R. Opher, M. Opher and A. Mann, Europhys. Lett.
\textbf{38} (1997) 245 (1997).

\bibitem{mann11} M. Revzen, R. Opher, M. Opher and A. Mann, J. Phys. A: Math
and Gen. \textbf{30} (1997) 7783.

\bibitem{mann2} M. Revzen and A. Mann, \emph{Casimir Effect - The Classical
Limit}, quant-ph/9803059.

\bibitem{mann3} J. Feinberg, A. Mann and M. Revzen, Ann. Phys.(NY) 288
(2001) 103.

\bibitem{brown} L. S. Brown and G. J. Maclay, Phys. Rev. \textbf{184} (1969)
1272.

\bibitem{robaschik} D. Robaschik, K. Scharnhorst and E. Wieczorek, Ann.
Phys.(N.Y.) \textbf{174} (1987) 401.

\bibitem{takagi} S. Tadaki and S. Takagi, Prog. Theor. Phys. \textbf{75 }
(1982) 262.

\bibitem{bag} A. Chodos, R. L. Jaffe, K. Johnson, C. B. Thorn and V. F.
Weisskopf, Phys. Rev. D \textbf{9} (1974) 3471.

\bibitem{saito1} K. Saito, Z. Phys. C \textbf{50} (1991) 69.

\bibitem{ravnd1} S. A. Gundersen and F. Ravndal, Ann. Phys. (N.Y.) \textbf{
182} (1988) 90.

\bibitem{ravnd2} C. A. L\"{u}tken and F. Ravndal, J. Phys. A: Math. Gen.
\textbf{21} (1988) L793.

\bibitem{ravnd3} C. A. L\"{u}tken and F. Ravndal, J. Phys. G: Nucl. Phys.
\textbf{10 }(1984) 123.

\bibitem{ravnd4} F. Ravndal and D. Tollefsen, Phys. Rev. D \textbf{40}
(1989) 4191.

\bibitem{svai3} R. D. M. De Paola, R. B. Rodrigues and N. F. Svaiter, Mod.
Phys. Lett. A \textbf{14} (1999) 2353.

\bibitem{eli1} E. Elizalde, F. C. Santos and A. C. Tort, Int. J. Mod. Phys.
A \textbf{18} (2003) 1761, hep-th/0206114.

\bibitem{johns1} K. Johnson, Acta Phys. Pol. B \textbf{6} (1975) 865.

\bibitem{ago1} L. A. Ferreira, A. H. Zimerman and J. R. Ruggiero, \emph{
Casimir Effect for Closed Cavities with Conducting and Permeable
Walls}, Instituto de F\'{\i}sica Te\'{o}rica, UNESP, Preprint
IFT-P-15/80.

\bibitem{ago11} W. Lukosz, Physica \textbf{56} (1971) 109

\bibitem{ago2} J. Ambj$\phi $rn and S. Wolfram, Ann. Phys. (N.Y.) \textbf{147
} (1983) 1.

\bibitem{ago3} J. Ambj$\phi $rn and S. Wolfram, Ann. Phys. (N.Y.) \textbf{147
} (1983) 33.

\bibitem{ago4} T. H. Boyer, Phys. Rev. A \textbf{9} (1974) 2078.

\bibitem{jura2} J. C. da Silva, A. Matos Neto, H. Queiroz Pl\'{a}cido, M.
Revzen and A. E. Santana, Physica A \textbf{292} (2001) 411.

\bibitem{sund1} P. Sundberg and R. L. Jaffe, Ann. Phys. (NY) \textbf{309}
(2004) 442.

\bibitem{fermion} C. Itzykson and J. B. Zuber, \emph{Quantum Field Theory}
(McGrow-Hill, New York, 1980).

\bibitem{sout1} K. Soutome, Z. Phys. C \textbf{40} (1988) 479.
\end{thebibliography}
\end{document}